# Concentration-dependent shear response of multi-chain amphiphilic block copolymer self-assemblies

Ehsan Kamali Ahangar, Dominic Robe, Elnaz Hajizadeh[*]
Soft Matter Informatics Research Group, Department of Mechanical Engineering, Faculty of Engineering and Information Technology, The University of Melbourne, Victoria 3010, Australia
[*]Corresponding author: ellie.hajizadeh@unimelb.edu.au

**Abstract**
Amphiphilic block copolymers self-assemble into diverse nanoscale morphologies with significant implications for drug delivery. This work presents systematic Brownian dynamics simulations of multi-chain diblock and triblock copolymers across dilute and semi-dilute unentangled regimes, hydrophobic fractions ($f$) of 0-1, and shear rates of 0-0.1 $ns^{-1}$. In the dilute regime, quiescent conditions yield spherical micelles evolving to cigar-like structures at $\dot{\gamma}\sim0.01$ $ns^{-1}$ and fragmenting at higher shear; varying $f$ produces dispersed chains ($f=0$), cigar-like ($f=0.25$), short cylindrical ($f=0.5$), and gnarled or worm-like ($f=0.75$) micelles, culminating in sheet-like phase-separated structures ($f=1$). While, in the semi-dilute regime, shear drives collective reorganisation toward sheet-like morphologies at moderate rates before fragmentation; the $f$-dependent progression yields cigar-like ($f=0.25$), sheet-like ($f=0.5$), and necklace micelles ($f=0.75$), with larger phase-separated domains at $f=1$. Triblock architectures form interconnected bridging networks in dilute conditions, achieving higher aspect ratios ($L_1/L_3\sim10-11$) at moderate shear, whereas in semi-dilute conditions, diblocks exhibit superior anisotropy through greater conformational flexibility. Rheological characterisation reveals a universal architectural inversion between equilibrium and flow conditions: diblocks show higher equilibrium viscosity while triblocks maintain superior viscosity under flow via bridging networks. Aggregation number scaling exponents of $\alpha=0.833$ in dilute, consistent with star-to-crew-cut bounds of 0.8 to 1.0, and $\alpha=1.07$ in semi-dilute confirm the concentration-driven transition between regimes. Viscoelastic analysis establishes universal non-terminal power-law scaling ($G'\sim\omega^{1.41\pm0.05}$, $G''\sim\omega^{0.635\pm0.45}$) across all conditions, governed by micellar relaxation dynamics independent of concentration or topology. These findings provide valuable insights into tailoring the injectability and flow behaviour of block copolymers in drug delivery formulations.

## 1. Introduction

Amphiphilic block copolymers, comprising covalently linked hydrophilic and hydrophobic segments, spontaneously self-assemble in selective solvents into a rich spectrum of nanoscale morphologies [1], [2], [3]. This organisation is thermodynamically driven by the minimisation of unfavourable contacts between the hydrophobic blocks and the solvent, a process analogous to the hydrophobic effect in small-molecule surfactant systems. Bates and Fredrickson [4] established the conceptual framework for block copolymer self-assembly, identifying the volume fraction of each block and the degree of chemical incompatibility between segments as the two primary variables governing phase behaviour. Depending on these variables along with concentration and solvent selectivity, block copolymers in solution form spherical micelles, cylindrical or wormlike micelles, vesicles, lamellar bilayers, bicontinuous networks, and more exotic structures including cigar-shaped, gnarled, and necklace micelles. Zhang and Eisenberg [5] demonstrated this morphological richness experimentally in polystyrene-b-poly(acrylic acid) solutions, obtaining spheres, rods, vesicles, and lamellae within a single polymer family simply by varying block ratio and solvent composition. The technological relevance of these morphologies is broad: polymeric micelles encapsulate hydrophobic therapeutics for tumour-targeted delivery; wormlike micellar networks impart desirable viscoelastic properties in personal care formulations; block copolymer membranes mediate selective ion transport in energy storage devices; self-assembled nanostructures template functional porous materials for catalysis and separation [6], [7]. A central concept for rationalising which morphology forms is the molecular packing geometry, captured by a balance between the hydrophilic weight fraction $f$, the interfacial curvature at the core–corona boundary, and the elastic stretching penalties of both blocks. Two well-established limiting regimes govern this balance. In the star-like micelle regime, the hydrophilic corona block greatly exceeds the hydrophobic core block in length, yielding assemblies that resemble star polymers with a small dense core enveloped by a thick, brush-like corona extending far into the solvent. Halperin and Alexander [8] showed that in this geometry, the free energy is dominated by the osmotic repulsion and elastic stretching of corona chains, with aggregate's curvature set by a balance of these contributions against interfacial tension. They also predicted that star-like micelles are predominantly spherical and kinetically robust. Zhulina and Borisov [9] extended this analysis to account for the crossover from star-like to crew-cut conditions, demonstrating analytically that as the core block lengthens relative to the corona, the preferred aggregate geometry shifts progressively from spheres through cylinders to bilayers due to decreasing interfacial curvature. In the crew-cut regime, a large dense core is capped by a thin, collapsed corona, and non-spherical morphologies become energetically accessible.

The equilibrium self-assembly of block copolymers has been extensively characterised as a function of molecular architecture and solution concentration using small-angle neutron scattering (SANS), small-angle X-ray scattering (SAXS), and oscillatory rheology. Pham et al. [10] demonstrated on associative triblock copolymers that storage and loss moduli and zero-shear viscosity are directly coupled to aggregation number, micellar radius, and intermicellar interactions quantified through structure factors, establishing hydrophobic end-block length as the principal handle for controlling gelation concentration and mechanical response. Tae et al. [11] extended this approach to fluoroalkyl-terminated poly(ethylene glycol) triblocks, resolving ordering transitions from disordered micellar solutions to face-centred cubic lattices through combined oscillatory rheology and SANS, and demonstrating that molecular design dictates the onset of long-range structural order. Garnier and Laschewsky [12] employed reversible addition fragmentation chain transfer polymerization to synthesize a series of amphiphilic diblock copolymers comprising poly(butyl acrylate) as the hydrophobic block paired with six chemically distinct hydrophilic blocks of varying hydrophilicity. In aqueous solution, the copolymers self-assembled into micellar aggregates with diameters ranging from 25 to 100 nm, with hydrophobic block length identified as the primary factor governing aggregate size. The micelles were highly stable upon dilution and exhibited dynamic chain exchange between populations, reflecting the non-frozen character of the poly(butyl acrylate) core. Polymers with longer hydrophobic blocks initially produced large transient aggregates that disappeared over time, indicating thermodynamically driven but kinetically slow micellization.

Particle-based simulations have provided direct molecular-level insight into equilibrium self-assembly. Liu et al. [13] used coarse-grained molecular dynamics simulations to study vesicle formation in BAB triblock copolymers in explicit aqueous solvent. Increasing B-block hydrophobicity accelerated vesiculation without altering the pathway, while excessive hydrophilicity suppressed vesicle formation in favour of multilayered spherical micelles. Yang et al. [14] employed dissipative particle dynamics (DPD) simulations to investigate the aqueous self-assembly of amphiphilic cyclic brush block copolymers. Six distinct self-assembled morphologies were identified (rods, plates, vesicles, large compound vesicles, bilayers, and spheres) depending on molecular composition and architecture. For vesicles, increasing solvophobic chain or backbone length reduced cavity size while thickening the membrane, with overall vesicle size remaining largely unchanged. Zhang et al. [15] employed Brownian dynamics simulations to investigate the self-assembly of amphiphilic diblock copolymers in dilute solution, systematically varying solvent quality, chain flexibility, block composition, and chain length. Poor solvent conditions for the hydrophobic blocks favoured single flower micelles, while decreasing chain flexibility promoted bridge structures connecting multiple hydrophobic cores. Increasing chain length drove a single-to-multi-flower micelle transition, and pre-assembled flower micelles could subsequently serve as building blocks for hierarchical assembly into well-defined multicompartment wormlike micelles upon further worsening of the solvent conditions. Using the Kremer-Grest FENE model [16], Hao et al. [17] employed Brownian dynamics simulations to study the single-chain conformation of an amphiphilic comb-like copolymer in dilute solution. They identified an optimal total chain length ($N^* \approx 220$) separating a unimolecular spherical micelle regime from a cylindrical micelle regime, with chain dimensions below $N^*$ following a power-law relationship whose exponent decreases with increasing side-chain length. In the dilute and semi-dilute unentangled regimes, Rouse [18] demonstrated analytically that bead-spring chains possess a discrete relaxation spectrum, and Brownian dynamics implementations of this framework correctly recover equilibrium chain-dimension and diffusion scaling, establishing the unentangled bead-spring chain as the natural computational baseline against which flow-induced departures from equilibrium can be quantified.

Beyond equilibrium structure, shear flow [47], [54], [55] can drive block copolymer assemblies far from their rest-state configuration, and understanding this response is directly relevant to processing and their industry applications. Cates and Candau [19] established the theoretical basis for the rheology of entangled wormlike micelles, demonstrating that the competition between reptation and chain-breaking timescales controls terminal relaxation and that in the fast-breaking limit a monoexponential decay emerges, characterised by a relaxation time equal to the geometric mean of the reptation and breaking times. Experimentally, McCauley et al. [20] revealed highly elastic behaviour, yield stress phenomena, and pronounced flow heterogeneity in triblock copolymer wormlike micellar solutions undergoing a period of strong elastic recoil and flow reversal after the onset of shear startup. In the semi-dilute unentangled regime, Dunstan and Harvie [21] applied shear-induced particle pressure theory, treating chains as compressible elastic blobs whose size is set by the balance between shear-induced interaction stress and elastic restoring forces and predicted progressive blob compression with increasing shear rate. Their rheo-optical measurements on poly(methyl methacrylate) solutions confirmed this compression quantitatively, establishing that shear measurably alters chain dimensions even without entanglement.

Non-equilibrium DPD and molecular dynamics simulations have substantially advanced the mechanistic understanding of flow-induced structural transitions. Sliozberg et al. [22] used DPD simulations to study how concentration and block architecture control the structure and viscoelastic properties of ABA triblock copolymers in midblock-selective solvent. They showed that the bridge fraction increases with both copolymer concentration and relative midblock volume, while shorter midblocks predominantly formed loops. Liu and Sureshkumar [23] demonstrated morphological memory in block copolymer nanovesicles under non-equilibrium [51], [52], [44], [45], [46], finding that above a critical strain amplitude the aggregate topology is permanently altered through a persistent inter-vesicle molecular bridge that survives extended relaxation after flow cessation. In the dilute and semi-dilute unentangled regimes, Colby et al. [24] developed a modified Rouse model predicting that chains adopt stretched Pincus blob conformations under shear, with Rouse modes exceeding the applied shear rate switching from dissipative to elastic behaviour and driving shear thinning quantitatively consistent with measurements on unentangled polystyrene melts and semi-dilute polyelectrolyte solutions. Prhashanna et al. [25] employed dissipative particle dynamics combined with the Lowe-Anderson thermostat (LA-DPD) to investigate the morphology and chain conformation of Pluronic triblock copolymer micelles in aqueous solution under simple shear [48], [49], [50] across a range of polymer volume fractions representing different concentrations or regimes. They reproduced concentration-dependent morphologies from spherical micelles at low concentrations through ellipsoidal and wormlike aggregates to a fragile gyroid-type percolating network at $\varphi = 0.5$, with all systems exhibiting shear thinning behaviour under flow. Contrary to the prevailing assumption of chains adopting exclusively loop or bridge conformations, chains in fact display a nearly uniform distribution of conformations at low shear rates, shifting markedly toward bridge conformations only under strong flows as rodlike micelles form.

Despite this substantial body of work, the flow-induced structural and rheological behaviour of block copolymer micellar systems in the dilute and semi-dilute unentangled regimes remains a critical and largely unexplored gap. This work addresses this gap through systematic Brownian dynamics (BD) simulations to establish a useful framework across a multi-dimensional parameter space. Above the critical overlap concentration, polymers are represented as elastic spheres with negligible excluded volume effects [26], [27], and entanglements are not included, assumptions consistent with the Rouse model which neglects both chain entanglement and hydrodynamic interactions [18], [28], [29]. For the first time, we systematically investigate: (i) how varying the hydrophobic fraction from 0 to 1 changes the accessible morphologies, spanning spherical, short cylindrical, cigar-shaped, sheet, gnarled or worm-like, and necklace micelles as well as sheet-like phase-separated structures, across concentrations from dilute to semi-dilute unentangled regimes; (ii) how diblock (AB) and triblock (ABA) architectures respond differently under identical solution conditions; and (iii) how imposed shear rates spanning from quasi-equilibrium to high-shear conditions drive morphological transitions. Structural characterisation is quantified through three metrics: the gyration radius of individual chains per micelle to characterise chain-level conformation and spatial extent (noting that overall aggregate size additionally depends on aggregation number, which is quantified separately through cluster analysis); aspect ratios derived from the eigenvalues of the gyration tensor ($L_1/L_3$, $L_1/L_2$, and $L_2/L_3$) to quantify structural anisotropy and flow-induced alignment; and cluster count to track aggregation state. These structural metrics are directly correlated with rheological properties including solution viscosity, viscoelasticity (storage modulus $G'$ and loss modulus $G''$), and micellar relaxation time ($\tau$) extracted from the inverse of the crossover frequency at which $G' = G''$. The results reported here across concentration, architecture, shear rate, and chemical composition provide direct design guidelines for engineering block copolymer drug carriers with precisely tailored nanoscale morphology, injectability, and flow behaviour under physiologically relevant conditions.

## 2. Methodology

### 2.1. Problem Definition and Coarse-Grained Model Block Copolymers

Figure 1 shows seven distinct micellar morphologies that emerge from multi-chain amphiphilic systems: (a) spherical micelle, (b) short cylindrical micelle, (c) cigar-like micelle, (d) sheet-like micelle, (e) gnarled or worm-like micelle, (f) necklace micelle, and (g) sheet-like phase-separated structure. Figure 2 presents the initial configurations of multi-chain diblock and triblock copolymers examined across a range of hydrophobic fractions with constant chain length $N=48$. The coarse-grained modelling approach examines systems containing 120 polymer chains ($N_{chain}=120$) across two concentration regimes. Figure 2a displays simulations at dilute concentration ($\emptyset=0.0719$), while Figure 2b shows the semi-dilute regime ($\emptyset=0.1816$). Both Figures show systematic variation of hydrophobic fraction ($f$) from completely hydrophilic ($f = 0$) through intermediate compositions ($f = 0.25, 0.5, 0.75$) to fully hydrophobic ($f = 1$) chains. We explore two architectural designs: diblock copolymers ($N_{block} = 2$) and triblock copolymers ($N_{block} = 3$). Figure 2c provides a schematic representation using shorter chains ($N = 12$) to clearly visualize the molecular composition, with blue beads representing hydrophobic segments (A) and yellow beads denoting hydrophilic segments (B). The block notation indicates the sequential arrangement along the polymer backbone, for example, ($A_3$-$B_6$-$A_3$) for the triblock architecture at $f = 0.5$. To investigate the influence of external flow on micelle formation and structural transitions, shear rates of $\gamma = 0, 0.003, 0.01, 0.03,$ and $0.1\ ns^{-1}$ are applied to the system, ranging from quiescent to progressively stronger flow conditions.

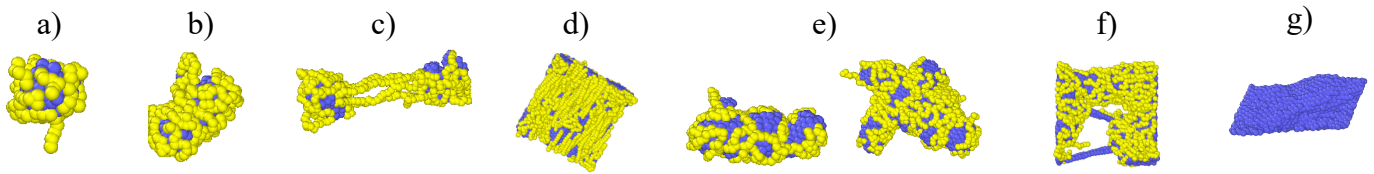

Figure 1: Different micellar structures of multi-chain block copolymers identified in this investigation: a) spherical micelle; b) short cylindrical micelle; c) cigar-like micelle; d) sheet-like micelle; e) gnarled or worm-like micelle; f) necklace micelle; and g) sheet-like phase-separated structure

$N_{chain} = 120, f = 0.5, N = 48,$ and $\emptyset = 0.0719$ (dilute concentration)

| $f = 0$ $N_{block} = 2\ or\ 3$ | $f = 0.25$ $N_{block} = 3$ | $f = 0.75$ $N_{block} = 2$ | $f = 1$ $N_{block} = 2\ or\ 3$ |
|---|---|---|---|

Figure (a)

$N_{chain} = 120, f = 0.5, N = 48,$ and $\emptyset = 0.1816$ (semi-dilute concentration)

| $f = 0$ $N_{block} = 2\ or\ 3$ | $f = 0.25$ $N_{block} = 3$ | $f = 0.75$ $N_{block} = 2$ | $f = 1$ $N_{block} = 2\ or\ 3$ |
|---|---|---|---|

Figure (b)

$N = 12$

| $f = 0$ | $f = 0.25$ | $f = 0.5$ | $f = 1$ |
|---|---|---|---|

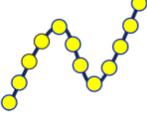

Figure 2: Initial configurations of multi-chain diblock and triblock copolymers examined across varying hydrophobic fractions with constant chain length $N$=48, at a) volume fraction of 0.0719 (dilute concentration); b) volume fraction of 0.1816 (semi-dilute concentration); c) schematic representation of the block structure containing 12 beads based on composition (hydrophobic beads depicted in blue and hydrophilic beads in yellow)

Non-bonded interactions between beads are modelled using the Lennard-Jones (LJ) potential [30]:

$$U_{LJ}(r_{ij}) = 4\epsilon\left[\left(\frac{\sigma}{r_{ij}}\right)^{12} - \left(\frac{\sigma}{r_{ij}}\right)^{6}\right] \quad (1)$$

The interaction strength ($\epsilon$) and size of bead ($\sigma$) change according to bead types. Three distinct interaction types are defined based on the chemical nature of the beads involved. For hydrophobic-hydrophobic interactions (type 2-2), parameters are set to $\epsilon_{22}$ = 4.195 *agr.nm²/ns²*, $\sigma$ = 1.5 *nm*, and $r_{cut}$ = 3.75 *nm*, ensuring robust attractive interactions enabling micelle core formation. The hydrophilic-hydrophilic interactions (type 3-3) employ weaker parameters: $\epsilon_{33}$ = 0.524 *agr.nm²/ns²*, $\sigma$ = 1.5 *nm*, and $r_{cut}$ = 1.68369 *nm*, which appropriately represent solvated polymer segments in aqueous environments. Cross-interactions at the micelle core-corona interface (type 2-3) utilize intermediate values: $\epsilon_{23}$ = 0.786 *agr.nm²/ns²*, $\sigma$ = 1.5 *nm*, and $r_{cut}$ as determined by the hydrophobic bead parameters.

Polymer chain connectivity is maintained through Finitely Extensible Nonlinear Elastic (FENE) potentials between consecutive bonded beads [16]:

$$U_{FENE}(r_{ij}) = -0.5\kappa_b R_m^2 \ln\left[1 - \left(\frac{r_{ij}}{R_m}\right)^2\right] \quad (2)$$

where the spring constant $\kappa_b$ and inter-bead separation $r_{ij}$ govern the bond elasticity. The maximum bond extension $R_m$ limits the separation $r_{ij}$ with $\kappa_b$=60 and $R_m$=2.25 nm. Each bond type (hydrophobic-hydrophobic, hydrophilic-hydrophilic, and hydrophobic-hydrophilic cross-bond interactions) incorporates the corresponding $\epsilon$ and $\sigma$ parameters from the LJ potential component.

### 2.2. Brownian Dynamics Simulation

The temporal evolution of bead positions in implicit solvent is captured using Brownian dynamics (BD) simulation technique. This approach efficiently incorporates solvent effects through friction and stochastic force components, enabling the computational study of large-scale self-assembly phenomena. Each bead's trajectory follows the equation of motion [15]:

$$r_i(t + \Delta t) = r_i(t) + v_i(t)\Delta t + \frac{\Delta t}{\gamma}(\boldsymbol{F}_i^C + \boldsymbol{F}_i^R) \quad (3)$$

Here, $r$ and $v$ denote position and velocity vectors for bead $i$, while $\boldsymbol{F}_i^C$ represents conservative forces arising from both bonded and non-bonded interactions. The stochastic force $\boldsymbol{F}_i^R$ satisfies the fluctuation-dissipation relation with friction coefficient $\gamma = k_B T/D$, where $D$ is the diffusion coefficient.

Implicit solvent effects are captured through the random force component, which obeys the fluctuation-dissipation theorem [15]:

$$\langle \boldsymbol{F}_i^R(t)\boldsymbol{F}_j^R(\acute{t})\rangle = -6\gamma k_B T \delta_{ij}\delta(t-\acute{t}) \qquad (4)$$

This coupling between friction and stochastic forces provides an effective thermostat. Random forces are generated using Gaussian random number generators to satisfy the fluctuation-dissipation theorem and maintain thermal equilibrium. The numerical integration employs a time step dt = 0.00005 $ns$ within the LAMMPS software framework, with Python-based post-processing scripts performing morphology-dependent analysis. In Equation 4, $k_B$, $T$, and $\delta$ denote the Boltzmann constant, temperature, and Dirac delta function $\delta(t-\acute{t})$ ensures temporal uncorrelation of random forces.

### 2.3. Simulation Procedure

Full details of the simulation framework are provided in the supplementary materials (Section S1), while model validation is reported in our previous work [31], where the Flory scaling exponents extracted from Brownian dynamics simulations were compared with those derived from experimental and theoretical studies under both good and poor solvent conditions. The generic coarse-grained bead-spring approach and its distinction from chemistry-specific methodologies are also discussed in detail. In the present study, the rheological response is examined across both dilute and semi-dilute concentration regimes, where shear-thinning behaviour is observed in both cases, in qualitative agreement with data reported in experimental studies and numerical investigations of block copolymer micellar solutions under comparable concentration regimes [32], [33].

## 3. Results and Discussion

### 3.1. Micellar Shape Analysis and Morphology of Block Copolymers

The purpose of this analysis is to identify and characterise the self-assembled morphologies that emerge across the multi-dimensional parameter space of concentration, molecular architecture, hydrophobic fraction, and shear rate (0-0.1 $ns^{-1}$), and to quantify how these morphologies deform under flow. All systems examined in this section consist of $N$ = 48 beads per chain, $N_{chain}$ = 120 chains, and a fixed hydrophobic fraction of $f$ = 0.5. The gyration tensor is computed for each chain within its micellar assembly, and its three principal components (eigenvalues $L_1$, $L_2$, and $L_3$ in descending order) are extracted, where $R_g^2 = L_1 + L_2 + L_3$. These eigenvalues represent the principal spatial dimensions of the chain: $L_1$ corresponds to the largest extent (aligned with the flow direction under shear), $L_2$ to the intermediate dimension, and $L_3$ to the smallest dimension. The ratios of these eigenvalues ($L_1/L_2$, $L_1/L_3$, and $L_2/L_3$) are calculated to quantify the overall shape anisotropy and flow-induced alignment of micellar assemblies, directly connecting eigenvalue distributions to the morphological categories. Figures 3a–3d present simulation snapshots systematically demonstrating the effects of shear rate and concentration regime on morphology for both diblock and triblock architectures, while Figures 3e and 3f quantify these morphological changes through the aspect ratios derived from the eigenvalues of the gyration tensor as a function of shear rate in the dilute and semi-dilute regimes, respectively. In the dilute regime (Ø = 0.0719), shown in Figures 3a and 3b, individual micelles exist as relatively isolated entities. At quiescent conditions ($\dot{\gamma} = 0$, leftmost column), both architectures form compact micellar structures where hydrophobic cores (blue) are shielded by hydrophilic coronas (yellow). However, a crucial architectural distinction emerges: Figure 3a shows that triblock copolymers at equilibrium establish interconnected networks between spherical micelles through bridging configurations where the central block spans between adjacent micellar cores, creating physical crosslinks. In contrast, Figure 3b shows diblock copolymers forming discrete spherical micelles without bridging. As shear rate increases to $\dot{\gamma} = 0.003$ $ns^{-1}$ (second column of Figure 3), both systems undergo orientational alignment, but with fundamentally different responses. Figure 3e reveals that triblock systems achieve higher aspect ratios (green, red, and purple curves) compared to diblock systems (blue, yellow, and black curves). This reflects the structural advantage of the bridging network: when the interconnected triblock network aligns with flow, the bridging chains between micelles become preferentially oriented, creating highly extended network structures where $L_1$ (along flow) increases dramatically as the network stretches collectively, while $L_2$ and $L_3$ (perpendicular dimensions) experience a moderate increase. The eigenvalue distribution becomes $L_1 \gg L_2 > L_3$, with $L_1/L_3$ ratios of 10-11. The visual appearance in Figure 3a at $\dot{\gamma} = 0.003$ $ns^{-1}$ shows this oriented, elongated network structure aligned with flow. In contrast, diblock micelles in Figure 3b at $\dot{\gamma} = 0.003$ $ns^{-1}$, as discrete units without bridging connections, align and elongate individually but to a lesser extent. The eigenvalue distribution shows moderate $L_1$ increase with $L_1/L_3 \sim 8$-9, creating elongated micelles but not the extreme network extension seen in triblocks. The absence of bridging limits the collective extension capability. Increasing shear rate to 0.01 $ns^{-1}$ causes both diblock and triblock polymers to get stretched (cigar-shaped micelle), but at high shear rates ($\dot{\gamma} = 0.03$-0.1 $ns^{-1}$) in Figure 3a, the interconnected network fragments under increased stress, breaking into more discrete micellar units [25]. The eigenvalue distribution transitions from the highly anisotropic network state ($L_1 \gg L_2 > L_3$) as bridges break: $L_1$ decreases (network contracts as connectivity is lost) while $L_3$ increases (individual units become less constrained), reducing overall aspect ratios.

In the semi-dilute regime ($\emptyset = 0.1816$), shown in Figures 3c and 3d, the physical behaviour changes fundamentally due to inter-micellar interactions, excluded volume effects, and corona overlap. Figure 3f reveals that aspect ratios that correlate with flow direction ($L_1/L_3$ and $L_1/L_2$) in semi-dilute conditions are systematically lower than dilute values (compared with Figure 3e) for non-equilibrium conditions under shear. To explain the reason behind this, in dilute conditions, individual micelles or small networks deform independently, allowing extreme elongation in one dimension (large $L_1$) while maintaining small perpendicular dimensions (small $L_2$, $L_3$). The structures adopt cigar-like or short cylinder-like morphologies with $L_1 \gg L_2 \approx L_3$, producing high $L_1/L_3$ ratios, and in semi-dilute conditions, the percolating network of corona overlap and transient contacts creates a fundamentally different deformation mode. Rather than individual elongation, the system undergoes collective reorganization, in which stress is distributed across multiple interconnected units. This promotes dimensional redistribution rather than pure elongation: the structures extend in two dimensions (flow and vorticity directions) while compressing in the velocity gradient direction, adopting sheet-like morphologies. Strikingly, Figures 3e and 3f reveal that in semi-dilute conditions, diblock systems achieve higher aspect ratios than triblock systems, particularly evident in $L_1/L_3$ and $L_1/L_2$ ratios. This represents a reversal from the dilute regime where triblocks have higher aspect ratios. This architectural reversal arises from the different response mechanisms to collective network reorganization, including (1) Diblock flexibility advantage: In the crowded semi-dilute environment, the linear A-B diblock architecture provides greater conformational flexibility for reorganization into highly anisotropic sheet structures. Diblock chains can more readily adopt extended planar configurations where chains lie flat and parallel within the sheet.

(2) Triblock structural constraints: The A-B-A triblock architecture with dual junction points creates topological constraints that resist extreme two-dimensional flattening. In the semi-dilute network, triblock chains cannot as easily adopt the planar configurations required for maximum sheet anisotropy. The bridging configurations that provided advantage in dilute conditions (enabling network extension) become a liability in semi-dilute conditions where they resist the dimensional redistribution required for sheet formation. (3) Network vs. micelle-dominated behaviour: In dilute conditions, triblock networks derive high aspect ratios from collective bridging extension. In semi-dilute conditions, both architectures form percolating networks, but the local chain architecture becomes determining. Diblock chains within the semi-dilute network can adopt more anisotropic local conformations, creating regions of high orientational order that propagate through the network. Triblock chains maintain more isotropic local configurations due to architectural constraints, resulting in networks with lower overall anisotropy.

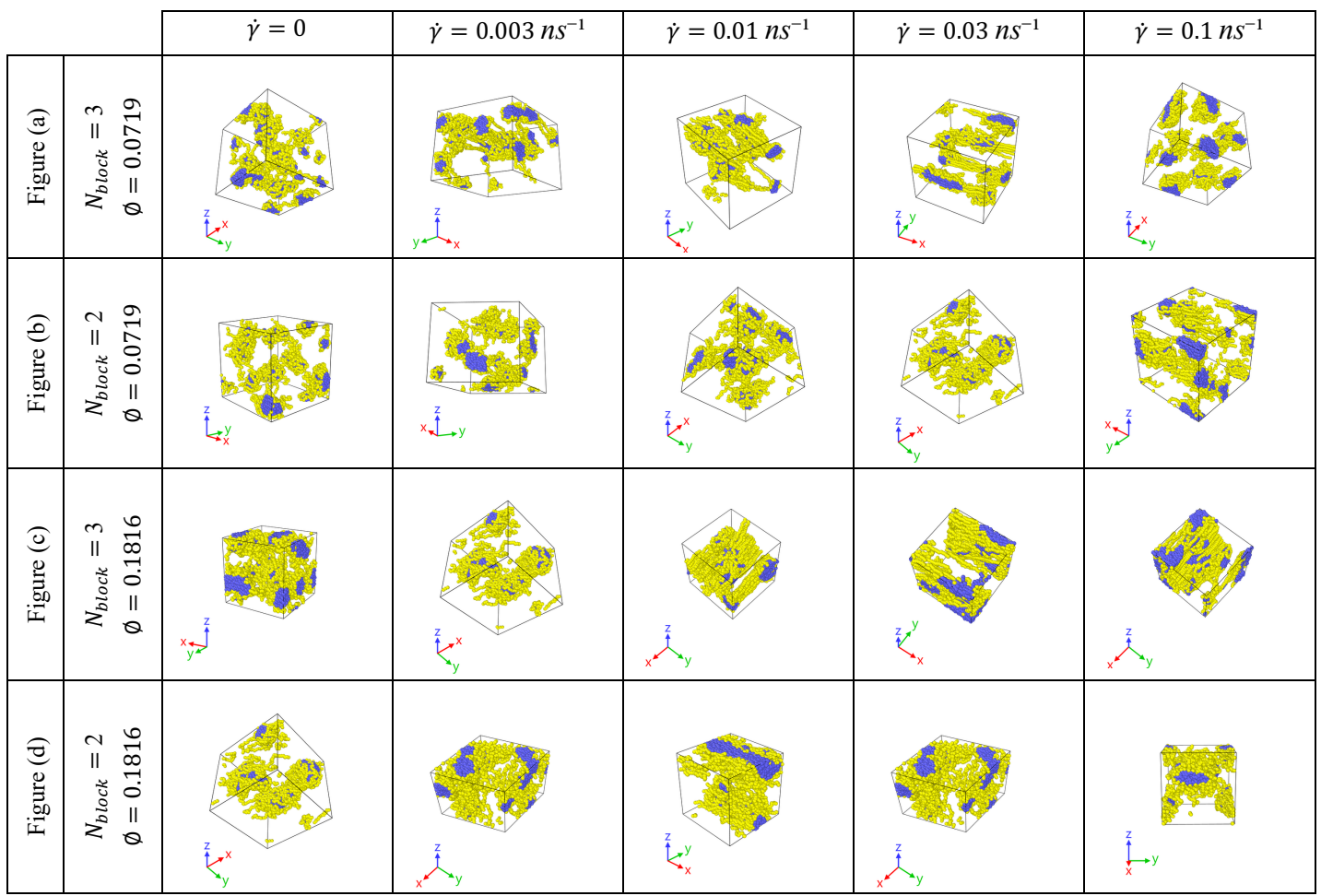

$N = 48, N_{chain} = 120$ and $f = 0.5$

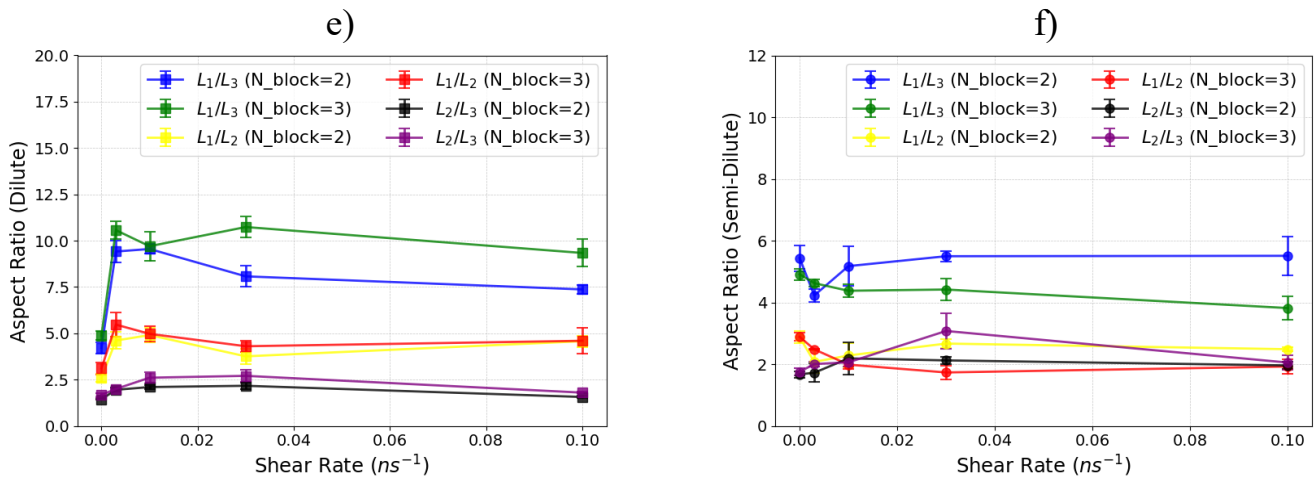

Figure 3: a-d) The effect of concenteration on chain shape of diblock and triblock copolymers at various shear rates at $N$=48, $N_{chain}$=120 and $f$=0.5 (∅ = 0.0719 dilute regime; ∅ = 0.1816 semi-dilute regime); e and f) aspect ratio derived from eigenvalues of $R_g$ tensor for block copolymers as a function of shear rate in dilute and semi-dilute regimes

Figure 4 systematically investigates the role of hydrophobic fraction in controlling molecular self-assembly and structural morphology at fixed shear rate ($\dot{\gamma}$ = 0.01 $ns^{-1}$) across both diblock and triblock architectures and concentration regimes with $N$=48 and $N_{chain}$=120. The hydrophobic fraction controls the fundamental self-assembly behaviour, creating three distinct regimes: individual chains ($f$ = 0), micellization ($f$ = 0.25, 0.5, 0.75), and phase separation ($f$ = 1). At $f$ = 0 (fully hydrophilic), leftmost column of Figure 4 shows individual polymer chains without any aggregation or micellization. Chains exist as extended random coils appearing as diffuse yellow networks throughout the simulation box. The complete absence of hydrophobic segments eliminates the thermodynamic driving force for self-assembly. Figures 4e and 4f show aspect ratios of ~5-40, reflecting flow-induced orientation of individual chains rather than organized structures. Within the micellization regime ($f$ = 0.25-0.75), the hydrophobic fraction controls the core-to-corona ratio, fundamentally determining micellar anisotropy and deformability. The micelle morphology varies systematically with $f$, controlled by the balance between flow-induced stress and interfacial tension.

At $f$ = 0.25 (second column of Figure 4), small hydrophobic cores (25%) and large hydrophilic coronas (75%) create anisotropic fiber-like or cigar-shaped micelles with aspect ratios $L_1/L_3$ ~ 10 for diblocks and ~15 for triblocks with eigenvalues $L_1 \gg L_2 > L_3$. Triblock architecture shows higher anisotropy due to bridging configurations between micelles creating more extended network structures. Minimal interfacial tension cannot resist flow-induced extensional stress, so the force ratio favours elongation along flow, with architectural differences most pronounced in this corona-dominated regime. At balanced composition ($f$ = 0.5), equal core-corona distribution (50%-50%) produces moderately elongated cylindrical structures. Aspect ratios $L_1/L_3$ ~ 10 with eigenvalues $L_1 > L_2 > L_3$, where architectural differences diminish as both diblock and triblock systems converge. The force ratio represents critical crossover where flow-induced stress and interfacial tension balance, transitioning from corona-dominated to core-dominated behaviour. At strong-core micelles ($f$ = 0.75), large cores (75%) and minimal coronas (25%) form gnarled or worm-like micelle. Aspect ratios $L_1/L_3$ ~ 8 with eigenvalues $L_1 \approx L_2 > L_3$ show minimal architectural effects as strong interfacial tension dominates. The force ratio favours compact shapes that resist flow-induced deformation. At $f$ = 1, complete absence of coronas leads to macroscopic phase separation. Aspect ratios $L_1/L_3$ ~ 5 with eigenvalues $L_1 > L_2 \approx L_3$ approach sheet-like geometry, where architectural differences vanish as thermodynamic surface minimization completely dominates across all molecular architectures. In Figure 4f, as mentioned in Figure 3 for shear-induced chains in semi-dilute regime (Ø = 0.1816), corona overlap and inter-micellar bridging create interconnected networks that undergo collective reorganization, producing systematically lower aspect ratios ($L_1/L_3$) than dilute conditions: ~10 for triblock copolymers at $f = 0.25$ (vs ~15 in dilute), ~4 at $f = 0.5$ (vs ~10 in dilute), and ~5 at $f = 0.75$ (vs ~8 in dilute). In semi-dilute regime shown in Figures 4c and 4d, networks deform from sheet-like ($f = 0.5$) to necklace ($f = 0.75$) geometries, exhibiting $L_1 \geq L_2 > L_3$ where $L_2$ grows substantially as structures expand laterally in the flow-vorticity plane, significantly reducing aspect ratios ($L_1/L_2$) despite larger overall structures. At $f = 1$, the higher concentration in the semi-dilute regime promotes formation of larger phase-separated domains as visualized in Figure 4c.

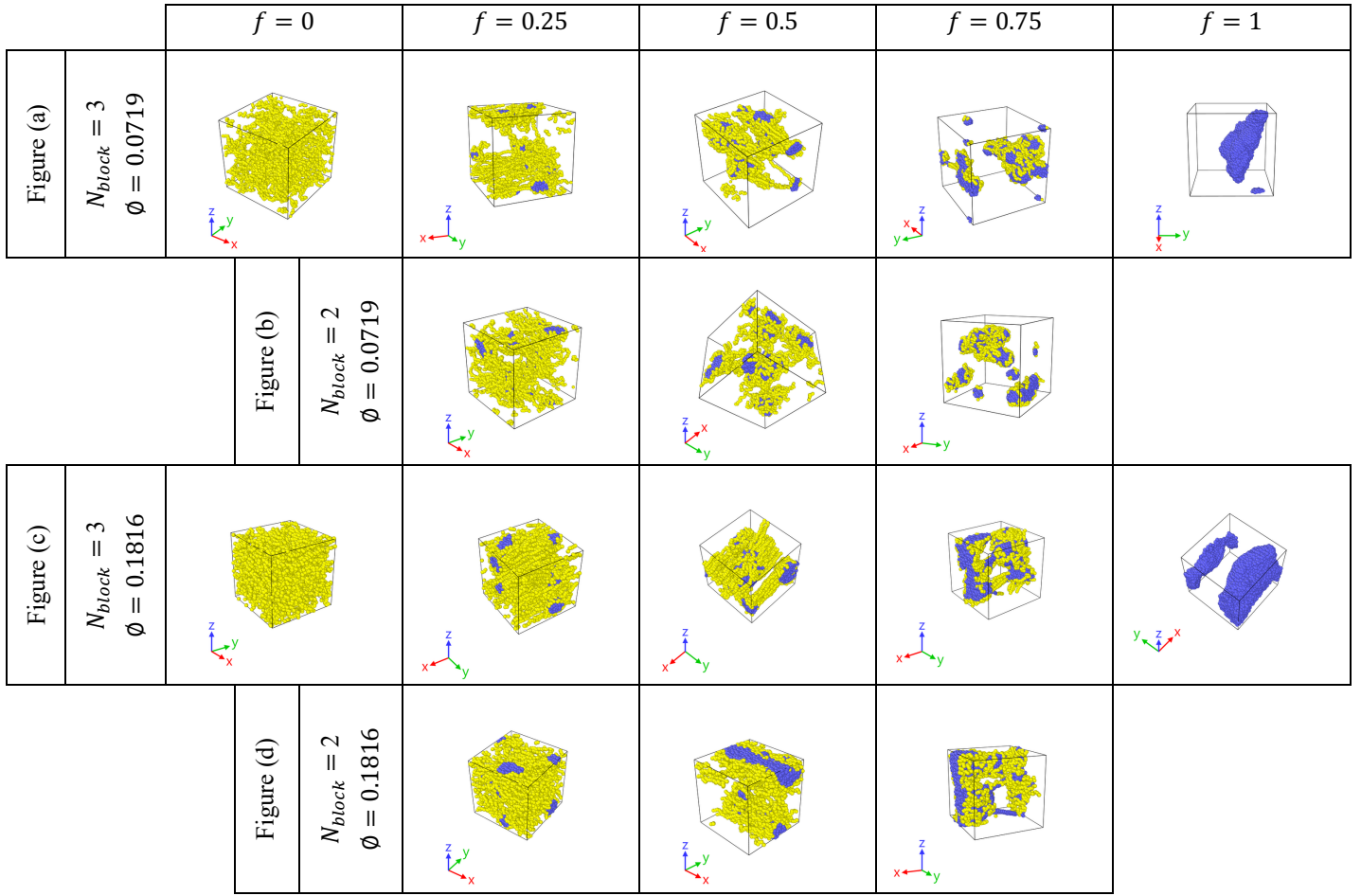

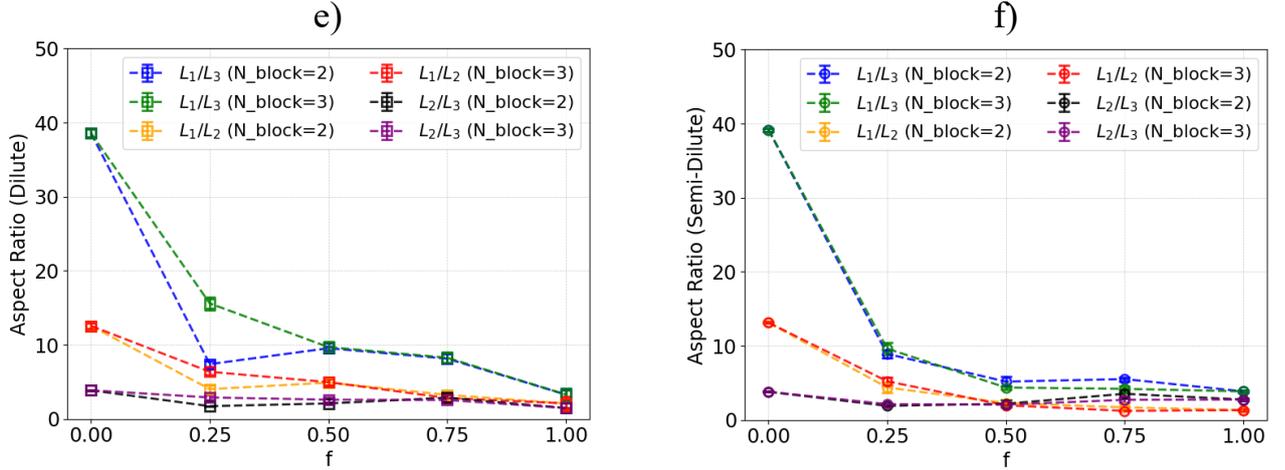

Figure 4: a-d) The impact of hydrophobic fraction on chain shape of block copolymers for two different molecular architectures (diblock and triblock) at $N=48$, $N_{chain}=120$ and $\dot{\gamma}=0.01$ $ns^{-1}$ ($\emptyset = 0.0719$ dilute regime; $\emptyset = 0.1816$ semi-dilute unentangled regime); e and f) aspect ratio extracted from eigenvalues of Rg tensor for block copolymers versus hydrophobic fraction in both dilute and semi-dilute regimes

### 3.2. Gyration Radius and Cluster Analysis

Figure 5 presents the size of individual diblock and triblock chains within each micelle and the number of identified micelles or phase separation domains under varying shear rates and chemical compositions at fixed $N=48$ and $N_{chain}=120$. The figure comprises four complementary subfigures that reveal the interplay between shear flow, polymer architecture, concentration regime, and hydrophobic content. Figure 5a displays the radius of gyration ($R_g$) as a function of shear rate for both diblock and triblock copolymers in dilute ($\varphi=0.0719$) and semi-dilute ($\varphi=0.1816$) regimes at $f=0.5$. All systems show initial increases from equilibrium ($\dot{\gamma} = 0$) to moderate shear rates, with a slight decrease at the higher shear rates (after $\dot{\gamma} \approx 0.01$ $ns^{-1}$) due to network fragmentation under high shear forces. The semi-dilute triblock system exhibits the largest normalized micelle size because polymer chains overlap significantly in the semi-dilute regime, creating inter-micellar bridges where the central block extends between different micellar cores (Figure 3c). The dilute diblock systems show lower normalized radii because micelles are spatially isolated with minimal inter-micellar interactions (Figure 3b). Within each concentration regime, triblocks display marginally larger structures than diblocks because the two hydrophobic end blocks can adopt extended conformations, whereas diblocks form simpler core-corona structures with one hydrophobic block anchored per chain. Figure 5b presents the cluster count as a function of shear rate. The general trend shows high cluster counts at equilibrium ($\dot{\gamma} = 0$), which decrease sharply to a minimum at critical shear rate ($\dot{\gamma} \approx 0.01$ $ns^{-1}$) as shear forces promote coalescence, followed by an increase at higher shear rates due to network breakdown and fragmentation. The dilute diblock system shows the highest count ($\approx 8$ micelles at equilibrium), while semi-dilute systems show the lowest counts ($\approx 1$-$2$ micelles) because extensive polymer overlap drives spontaneous aggregation into fewer, larger structures (Figures 3c-d and 5e). The semi-dilute triblock exhibits the most dramatic reduction with shear ($\approx 4$ to $1$-$2$ micelles) because shear facilitates inter-micellar connections through mobile central blocks.

In the dilute regime, the mean aggregation number is extracted as $\langle n \rangle = \frac{N_{chain}}{cluster\ count} = 120/8.5 \approx 14$ chains per micelle at equilibrium. With the hydrophobic block length fixed at $N_B = f \times N = 0.5 \times 48 = 24$ beads, the scaling exponent is extracted as $\alpha_{Dilute} = \frac{\ln(\langle n \rangle_S)}{\ln(N_B)} = \frac{\ln(14)}{\ln(24)} = \frac{2.639}{3.178} = 0.833$. This value falls within the theoretical bounds of [0.8, 1.0] established by Halperin–Alexander (star-like limit, α = 0.80 and crew-cut limit, α = 1.0) [8], validating our simulations against established dilute solution theory. In the semi-dilute unentangled regime, applying the identical extraction method with the same $N_B = 24$, the mean aggregation number increases to $\langle n \rangle = 120/4 = 30$ chains per micelle, yielding: $\alpha_{Semi-Dilute} = \frac{\ln(\langle n \rangle_{SD})}{\ln(N_B)} = \frac{\ln(30)}{\ln(24)} = \frac{3.401}{3.178} = 1.07$. This effective exponent exceeds the dilute upper bound of 1.0. Since the extraction method, chain length, and composition are identical across both regimes, the only quantities that change are the cluster count (from 8.5 to 4) and consequently $\langle n \rangle$ (from 14 to 30). This departure from dilute scaling limits is a direct reflection of the physical effect of concentration on aggregation. It is consistent with coronal overlap and screening effects at elevated concentration driving stronger N-dependence of aggregation number, confirming that the system has transitioned into the semi-dilute unentangled regime.

Figure 5c examines radius of gyration versus hydrophobic fraction at $\gamma = 0.01$ $ns^{-1}$. Dilute and semi-dilute systems exhibit non-monotonic and monotonic behaviours, respectively. Critically, the dilute systems show a clear decrease in $R_g$ at *f*=1.0 compared to *f*=0.75, while semi-dilute systems maintain elevated $R_g$ values with only a slight decrease at *f*=1.0 (notably, $R_g$ at *f*=1.0 remains significantly higher than at *f*=0 for semi-dilute systems). The semi-dilute triblock system shows the most pronounced variation, with $R_g$ increasing from ≈ 9 *nm* at *f*=0.25 to nearly 17 *nm* at *f*=0.75-1.0 (Figures 4c and 5f) because high hydrophobic fraction combined with chain overlap enables extensive network formation, in which the bridging architecture connects multiple cores into sprawling aggregates. Semi-dilute diblocks show more modest increases (Figure 4d) because each chain contributes to only one micelle core, limiting aggregate interconnectivity. The dilute systems show modest variations (Figure 4a) because the absence of inter-micellar overlap confines size changes to single-micelle properties. The decrease at *f*=1.0 in dilute systems occurs because complete absence of hydrophilic blocks leads to compact, collapsed structures that minimize solvent exposure. In contrast, semi-dilute systems maintain extensive networks even at *f*=1.0, resulting in a larger chain size due to smaller applied volume fraction (sheet-like morphology, as visualized in Figure 4c). Figure 5d shows cluster count variations with hydrophobic fraction. All systems show a decreasing trend with increasing *f*, exhibiting the highest counts at *f*=0.25 and converging dramatically to approximately one micelle or single phase-separated domain at *f*=1.0 as the enthalpic penalty for hydrophobic-solvent contact grows, promoting fusion to minimize interfacial area. The dilute diblock system shows the highest initial count (≈ 7 micelles at *f*=0.25), while semi-dilute systems begin with lower counts (≈ 4 micelles) since chain overlap already promotes aggregation even at moderate *f* (Figure 4c-d). At *f*=1.0, the overwhelming hydrophobic character dominates assembly, effectively obliterating the influence of concentration regime.

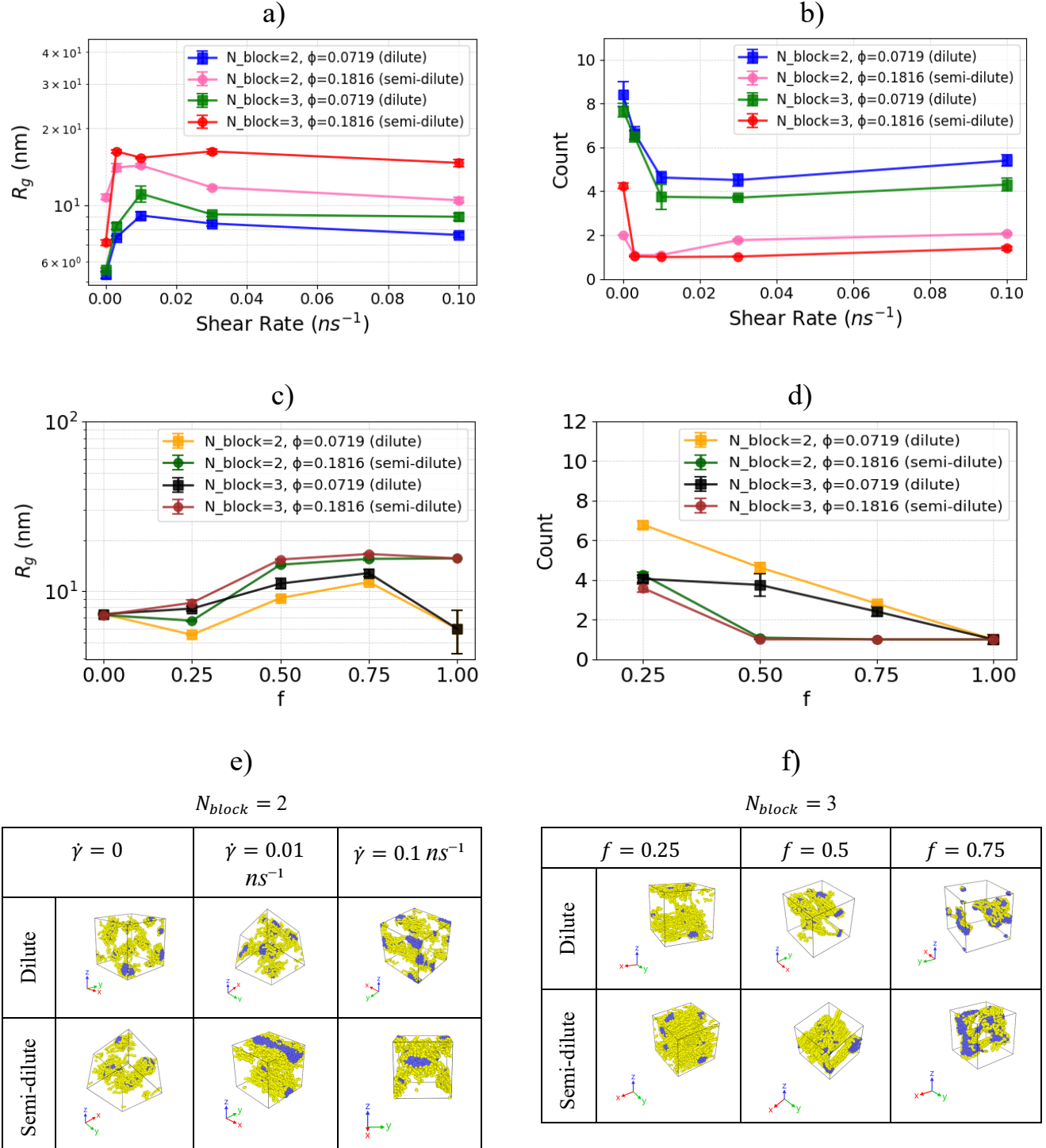

Figure 5: Radius of gyration (a, c) of individual chains within each micelle and individual cluster count (b, d) for diblock and triblock copolymers as a function of shear rate at $N=48$, $N_{chain}=120$ and $f=0.5$ (a, b); hydrophobic fraction (c, d) in dilute and semi-dilute regimes at $N=48$, $N_{chain}=120$ and $\dot{\gamma}=0.01$ $ns^{-1}$; e) snapshots of diblock copolymers at different shear rates ; and f) hydrophobic fraction-dependent snapshots of triblock copolymers at fixed $\dot{\gamma}=0.01$ $ns^{-1}$

### 3.3. Solution Viscosity

The purpose of this analysis is to establish how the self-assembled structures identified in Sections 3.1 and 3.2 translate into macroscopic rheological behaviour. Specifically, we aim to determine how molecular architecture, concentration regime, and chemical composition collectively control the viscosity response under shear. Characterisation is performed at the bulk solution level through the solution viscosity ($\eta$), calculated using the Green-Kubo relation. At zero shear rate, $\eta$ is obtained from three independent off-diagonal components of the stress autocorrelation function as $\eta=(V/k_BT)\int_0^\infty \langle \sigma_{\alpha\beta}(0).\sigma_{\alpha\beta}(t)\rangle$ dt [34], where averaging over three components (xy, xz, yz) enhances statistical accuracy. For non-zero shear rates, the stress autocorrelation function in the shear direction is divided by the applied shear rate to yield the shear viscosity, directly connecting the stress response of the micellar microstructure to macroscopic flow resistance. Figure 6a displays solution viscosity versus shear rate at $N=48$, $N_{chain}=120$ and $f=0.5$, revealing distinct rheological behaviours for each system. All systems exhibit shear-thinning behaviour, in which viscosity decreases with increasing shear rate [25]; however, the magnitude and rate of viscosity reduction vary dramatically based on molecular architecture and concentration regime. A notable architectural inversion occurs when comparing equilibrium and flow conditions. At equilibrium ($\dot{\gamma}=0$), diblock systems exhibit higher viscosity than triblocks in both concentration regimes: semi-dilute diblock (pink line, $\eta \approx 120$ *Pa.s*) exceeds semi-dilute triblock (red line, $\eta \approx 100$ *Pa.s*), and dilute diblock (blue line, $\eta \approx 40$ *Pa.s*) slightly exceeds dilute triblock (green line, $\eta \approx 30$ *Pa.s*). This occurs because diblock structures exhibit slower structural relaxation dynamics due to their linear architecture and restricted conformational rearrangement, while triblock bridging connections can rapidly exchange and reform, allowing faster relaxation and lower viscosity at rest. Under shear ($\dot{\gamma} > 0$), this viscosity ranking completely reverses, and triblock systems maintain higher viscosity than diblocks across all shear rates. Under low shear, triblock bridging configurations become locked and oriented by flow as shown in Figures 3 and 6c, creating stable physical crosslinks that resist breakdown, while diblock structures experience rapid disruption as flow alignment breaks apart their less stable aggregates. The viscosity of semi-dilute triblock decreases gradually to $\eta \approx 1$ *Pa.s* at $\dot{\gamma} = 0.1$ $ns^{-1}$, while the semi-dilute diblock drops sharply to the same value. This behaviour arises because bridging in triblock systems forms a more interconnected structure between micellar domains, resulting in greater resistance to flow.

Figure 6b examines solution viscosity versus chemical composition at $N=48$, $N_{chain}=120$ and $\dot{\gamma}=0.01$ $ns^{-1}$. All systems overall show increasing viscosity with increasing $f$, but the magnitude differs dramatically between concentration regimes and architectures. Across the entire range of $f$, semi-dilute systems exhibit significantly higher viscosity than dilute systems due to chain overlap enabling network formation (leading to higher $R_g$) and greater resistance to flow, as visualized in Figures 3, 4, and 6d. The semi-dilute triblock system (red line) exhibits the most dramatic increase, rising from $\eta \approx 0.4$ *Pa.s* at $f=0$ to approximately 5-7 *Pa.s* at $f=0.5$-0.75, maintaining this value at $f=1.0$, reflecting structural evolution from weak to extensively interconnected networks. The semi-dilute diblock system (green line) follows a similar trend with more modest magnitude, increasing to $\eta \approx 5$ *Pa.s* at $f=0.75$, because architectural constraints limit network connectivity. The dilute triblock system (black line) shows moderate increase from $\eta \approx 0.15$ *Pa.s* at $f=0$ to 1 *Pa.s* at $f=0.75$. The dilute diblock system (yellow line) exhibits the smallest variation, reaching $\eta \approx 1$ *Pa.s* at $f=0.75$. The viscosity data for both architectures at $f=0.5$ and $f=0.75$ reveal distinct composition-dependent behaviour. For diblock copolymers, which possess a single hydrophobic end block, decreasing hydrophilic block length promotes the formation of denser, isolated hydrophobic cores, resulting in greater resistance to flow. In contrast, triblock copolymers, which possess two hydrophobic end blocks, exhibit reduced resistance to flow as the hydrophilic midblock length decreases below the threshold required to bridge neighbouring micellar aggregates, weakening the transient network connectivity. Viscosity is controlled not simply by aggregate number but by connectivity and deformability. Semi-dilute triblocks at $f=1.0$ form extensive networks (sheet-like structure as shown in Figure 4c), while dilute diblocks or triblocks at $f$=1.0 form compact separated domains with smaller $R_g$ resulting in lower viscosity. The combination of bridging architecture, concentration regime enabling overlap, and high hydrophobic fraction is necessary to achieve maximum viscosity enhancement (visible at $f$=0.5 for triblock), while at extreme compositions ($f$=0 or $f$=1), chemical structure dominates.

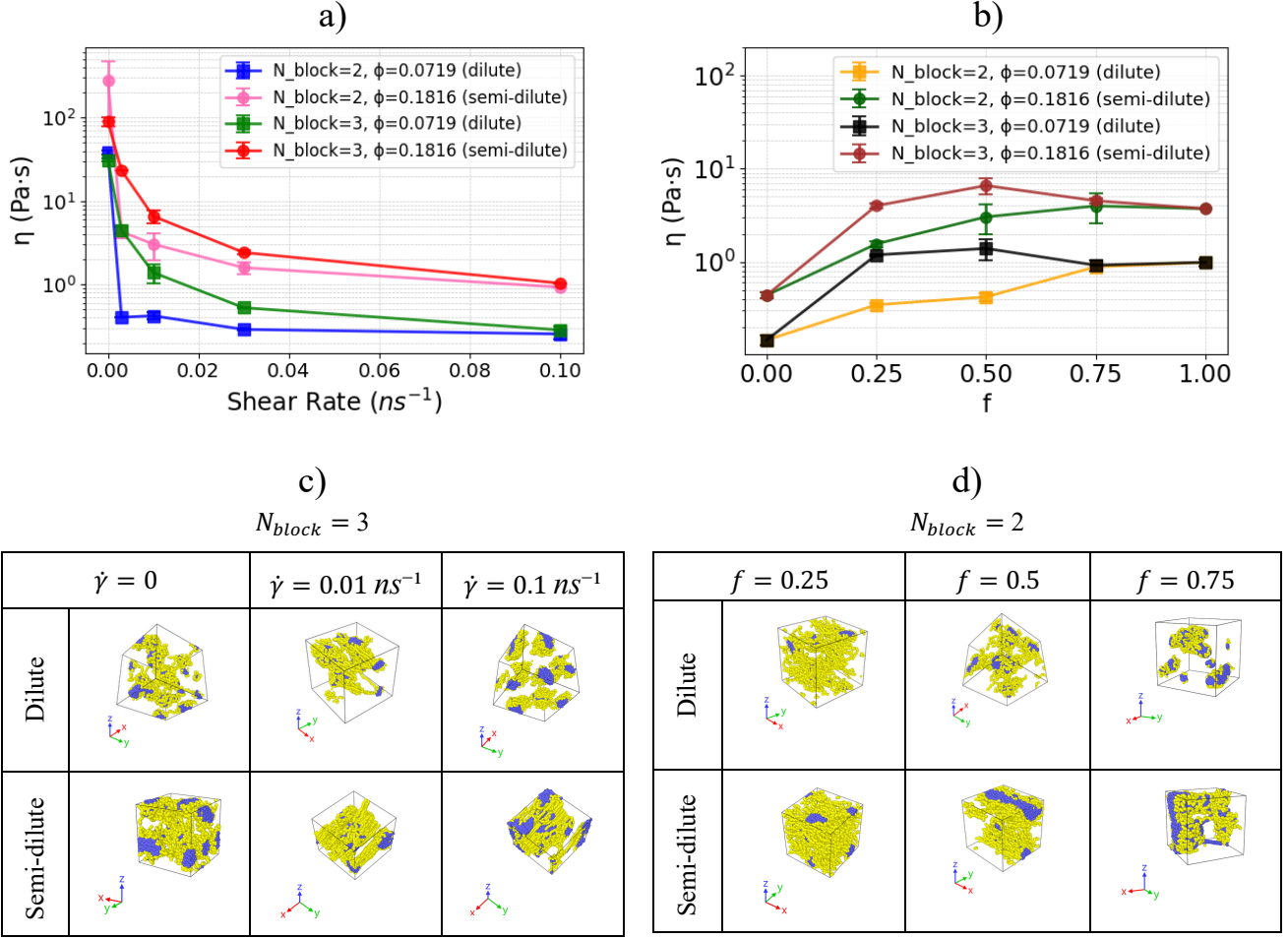

Figure 6: Solution viscosity (η) of diblock and triblock copolymers as a function of a) shear rate at $N$=48, $N_{chain}$=120 and $f$=0.5; b) hydrophobic fraction at volume fractions of 0.0719 (dilute regime) and 0.1816 (semi-dilute regime) with $N$=48, $N_{chain}$=120 and $\dot{\gamma}$=0.01 $ns^{-1}$; c) snapshots of triblock copolymers at zero, middle and high shear rates in both regimes; and d) composition-dependent snapshots for diblock copolymers at shear rate of 0.01 $ns^{-1}$ in both regimes

### 3.4. Viscoelasticity and Micellar Relaxation Time

The viscoelastic response of diblock and triblock copolymer systems is characterised through the frequency-dependent storage modulus (G′) and loss modulus (G″). These are derived from the stress relaxation modulus G(t), which is calculated using the Green-Kubo relation. For quiescent conditions, G(t) is obtained from $G(t) = \frac{V}{k_B T} \langle \sigma_{\alpha\beta}(0) \cdot \sigma_{\alpha\beta}(t) \rangle$, where averaging is performed over three independent off-diagonal components of the stress autocorrelation function, and for non-zero shear rates, the stress autocorrelation function in the shear direction is divided by the applied shear rate. The frequency-dependent moduli G′(ω) and G″(ω) are then obtained from G(t) through Fourier transformation [34]:

$$G'(\omega) = \omega \int_0^\infty G(t)\sin(\omega t)\, dt \qquad (5)$$

$$G''(\omega) = \omega \int_0^\infty G(t)\cos(\omega t)\, dt \qquad (6)$$

The micellar relaxation time is extracted as $\tau = 1/\omega_c$ from the crossover frequency $\omega_c$ at which $G' = G''$, characterising the timescale over which the micellar solution fully relaxes its structure through collective formation and dissolution events.

Figures 7a and 7b present the frequency-dependent viscoelastic spectra for diblock and triblock copolymers at $N = 48$ and $f = 0.5$ in the dilute regime. Across the low-frequency range, $G''$ exceeds $G'$ for both architectures, establishing that energy dissipation governs the long-timescale response. As frequency increases, $G'$ rises more steeply than $G''$ in both cases, eventually driving a crossover to elasticity-dominated behaviour at high frequencies. This crossover pattern is a direct signature of the micellar microstructure: at timescales longer than the micellar relaxation time, the assemblies have sufficient time to fully reorganise, and the system behaves as a viscous fluid, whereas at shorter timescales, the transient micellar network resists deformation elastically before reorganisation can occur. A clear distinction between the two architectures emerges when comparing Figures 7a and 7b: triblock copolymers develop higher $G'$ values at high frequencies relative to diblocks under equivalent conditions. This elevated elastic response originates from the bridging network visible in Figure 3a, where individual triblock chains simultaneously anchor into two spatially distinct micellar cores and the hydrophilic middle block spans the intervening space. These bridging configurations act as elastic connectors between neighbouring cores, storing elastic energy under deformation and raising $G'$ above the values attainable by diblock systems, where chains contribute to only one micellar core and no such inter-micellar elastic storage mechanism exists, consistent with the discrete micellar organisation in Figure 3b.

Figures 7c and 7d present the composition-dependent viscoelastic spectra for diblock and triblock copolymers at fixed $N = 48$ and $\dot{\gamma} = 0.01\ ns^{-1}$ in the dilute regime. At $f = 0$, purely hydrophilic chains exhibit lowest moduli from low to moderate frequencies, consistent with the complete absence of self-assembled structure confirmed by the dispersed chain morphologies in Figures 4a and 4b. The introduction of hydrophobic content at $f = 0.25$ triggers micellisation, with both $G'$ and $G''$ rising substantially as the cigar-shaped assemblies identified in Figures 4a and 4b develop under shear. For diblock copolymers (Figure 7c), both moduli increase progressively with $f$ but $G''$ continues to exceed $G'$ across most of the frequency ranges studied at all compositions, reflecting the fundamentally dissipative character of the discrete micellar architecture where each chain contributes to only one core. At $f = 1.0$, both moduli rise further due to steric packing constraints within the compact phase-separated domains visible in Figure 4b, but the absence of a well-defined core-corona interface eliminates organised micellar exchange dynamics. For triblock copolymers (Figure 7d), the compositional evolution is qualitatively distinct. At $f = 0.25$, a more pronounced high-frequency $G'$ contribution compared to the equivalent diblock case reflects the early establishment of bridging configurations between the cigar-shaped assemblies visible in Figure 4a, introducing inter-micellar elastic storage absent in diblock systems. The most significant spectral change occurs between $f = 0.5$ and $f = 0.75$, in which $G'$ develops a substantially stronger plateau extending to lower frequencies, tracking the topological transition toward fewer but larger and more stable assemblies. This is confirmed by the decrease in cluster count from ≈4 to ≈2 in Figure 5d and the morphological transition toward worm-like structures in Figure 4a.

Figures 7e and 7f present the viscoelastic spectra for diblock and triblock copolymers in the semi-dilute regime at fixed $N = 48$ and $f = 0.5$. The most prominent observation when comparing Figures 7e and 7f with their dilute counterparts in Figures 7a and 7b is the systematic increase in the absolute values of both G′ and G″ across the entire frequency range. This elevation in moduli directly reflects the higher polymer concentration in the semi-dilute regime, where chain overlap promotes the formation of more extensively interconnected assemblies, as confirmed by the lower cluster counts and larger $R_g$ values in Figures 5a and 5b. The same qualitative trends with shear rate observed in the dilute regime are preserved, with the semi-dilute triblock system (Figure 7f) maintaining consistently higher G′ values than the semi-dilute diblock system (Figure 7e) at equivalent shear rates, reflecting the additional elastic contribution of bridging configurations within the more densely interconnected network visible in Figure 3c. Figures 7g and 7h present the composition-dependent viscoelastic spectra for diblock and triblock copolymers at fixed $N = 48$ and $\dot\gamma = 0.01$ $ns^{-1}$ in the semi-dilute regime. Unlike the shear rate-dependent spectra in Figures 7e and 7f, varying $f$ fundamentally alters the nature of the self-assembled structures, producing a richer and more architecturally differentiated viscoelastic response, where the progressive strengthening of G′ with f directly tracks the structural evolution from cigar-shaped toward sheet-like and necklace morphologies confirmed by Figures 4c and 4d for triblock and diblock copolymer systems, respectively.

From Figures 7a–7h, both G′ and G″ exhibit power-law scaling deviating markedly from terminal flow behaviour (G′ ~ $\omega^2$, G″ ~ $\omega^1$) across all conditions, establishing a universal non-terminal scaling of G′ ~ $\omega^{1.41\pm0.05}$ and G″ ~ $\omega^{0.635\pm0.45}$ applicable across both dilute and semi-dilute unentangled regimes. This convergence arises because scaling behaviour is governed by the intrinsic dynamics of micellar assemblies rather than inter-micellar interactions or concentration-dependent topology. Grounded in the frameworks of Halperin and Alexander [8] and Kumar et al. [35], the timescale ratio between fast single-chain expulsion and slow collective reorganisation depends on molecular parameters fixed at constant $N$ and $f$ regardless of concentration, while consistent with the Rouse model for unentangled chains, concentration affects only the absolute magnitude of moduli rather than their frequency dependence. Although numerically similar between architectures, the mechanisms responsible for non-terminal behaviour are fundamentally different. For diblocks, deviation arises solely from polydispersity of discrete micellar sizes, which generates a broad distribution of single-chain exchange timescales, combined with the coexistence of fast and slow relaxation processes. For triblock copolymers, the ABA architecture introduces an additional cooperative pathway where complete junction relaxation demands sequential dissociation of both anchored end blocks from their respective cores, a constraint absent in diblocks that broadens the relaxation spectrum further and sustains higher elastic storage at high frequencies, with this effect amplified in the semi-dilute regime through greater bridging junction density; though this does not alter the underlying scaling exponents.

In Figure 8a, dilute diblock copolymers exhibit a non-monotonic response, i.e., micellar relaxation time ($\tau$) decreases from its highest value at weak shear to a minimum at moderate shear, then recovers at $\dot\gamma = 0.1$ $ns^{-1}$. The initial decrease reflects flow-assisted reorganisation, in which shear transiently exposes hydrophobic segments at the core-corona interface, while the recovery at high shear arises from flow-induced ordering that reduces effective inter-micellar encounters. Semi-dilute diblocks follow the same non-monotonic trend but with systematically lower $\tau$, reflecting closer core proximity from chain overlap providing more frequent reorganisation pathways. Dilute triblock copolymers show the opposite behaviour, where $\tau$ increases dramatically and monotonically from its lowest value at weak shear, reflecting the topological advantage of bridging networks visible in Figure 3a, in which dissociated ends remain tethered and rapidly locate neighbouring cores without bulk diffusion. As shear dismantles bridging connections confirmed by increasing cluster count in Figure 5b, this geometric confinement advantage is lost, thereby progressively raising τ. Semi-dilute triblocks follow the same increasing trend but with consistently higher τ, reflecting greater flow-induced locking of bridging configurations under mechanical tension. Importantly, at γ̇ = 0.1 $ns^{-1}$, τ converges toward a common value for all systems, indicating that at sufficiently high shear, flow-induced fragmentation dominates regardless of architecture and concentration regime.

In Figure 8b, dilute diblock systems maintain constant relaxation time since each chain contributes a single hydrophobic block anchored into one discrete micellar core, making the rate-limiting step for collective reorganisation governed by the energetics of crossing the core-corona interface rather than the depth of burial within the core. Despite changes in core-to-corona ratio as *f* increases from 0.25 to 0.75, the activation energy governing interfacial crossing remains approximately constant under the applied shear rate of 0.01 $ns^{-1}$, keeping the reorganisation dynamics insensitive to chemical composition. Dilute triblocks show the lowest τ across all compositions but with an increasing trend with *f*, reflecting the progressive elimination of bridging pathways as assemblies consolidate into fewer larger structures confirmed by the decreasing cluster count in Figure 5d, gradually shifting reorganisation from geometrically facilitated toward increasingly constrained collective events. Both semi-dilute diblock and triblock systems show increasing τ with *f* and maintain similarly elevated τ values across all compositions compared to dilute triblocks. This behaviour arises because chain overlap at higher concentration creates comparable inter-micellar connectivity for both architectures, partially equalising reorganisation pathways and producing a shared increasing trend driven by progressive network consolidation with growing hydrophobic content, regardless of architectural differences.

**Dilute Regime**

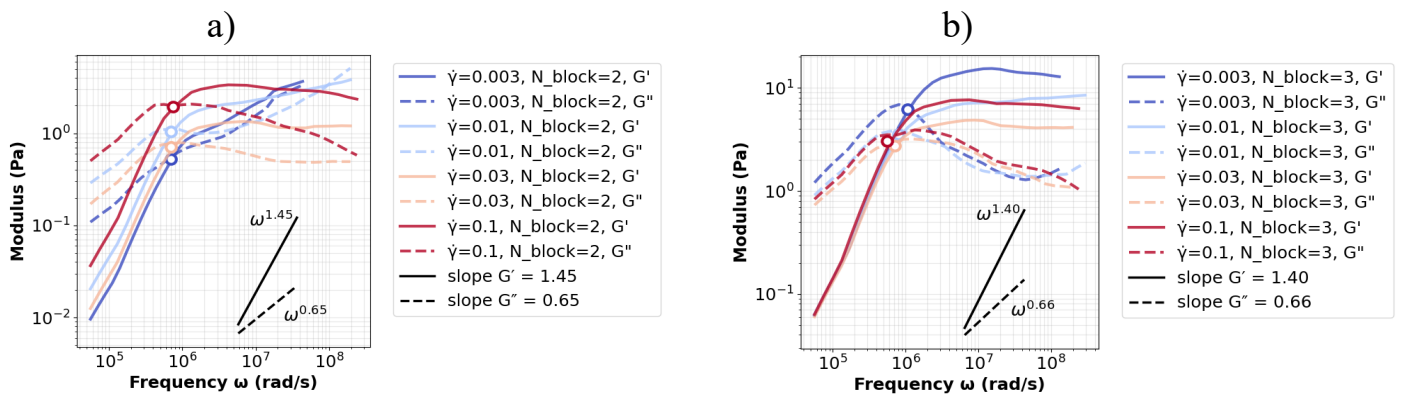

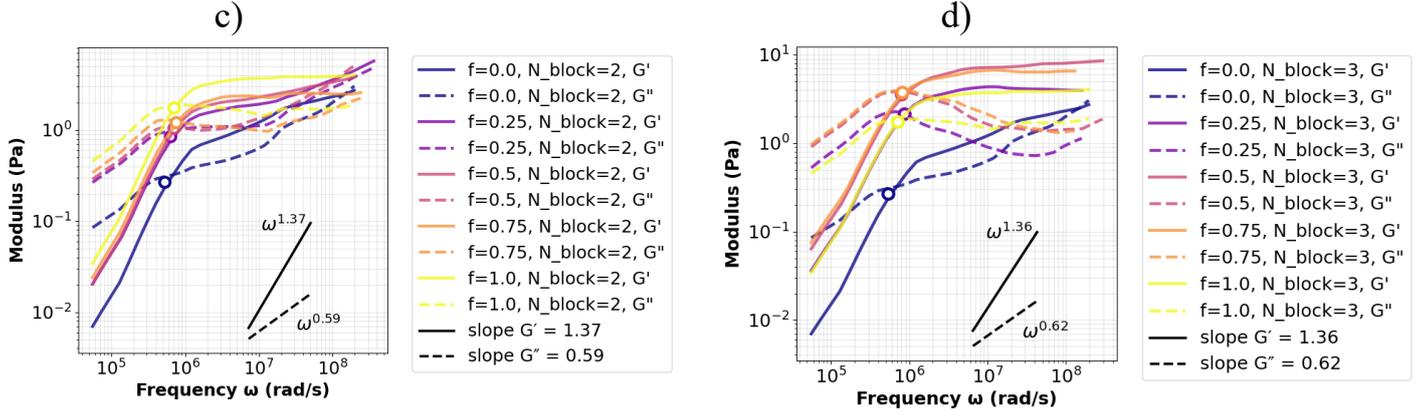
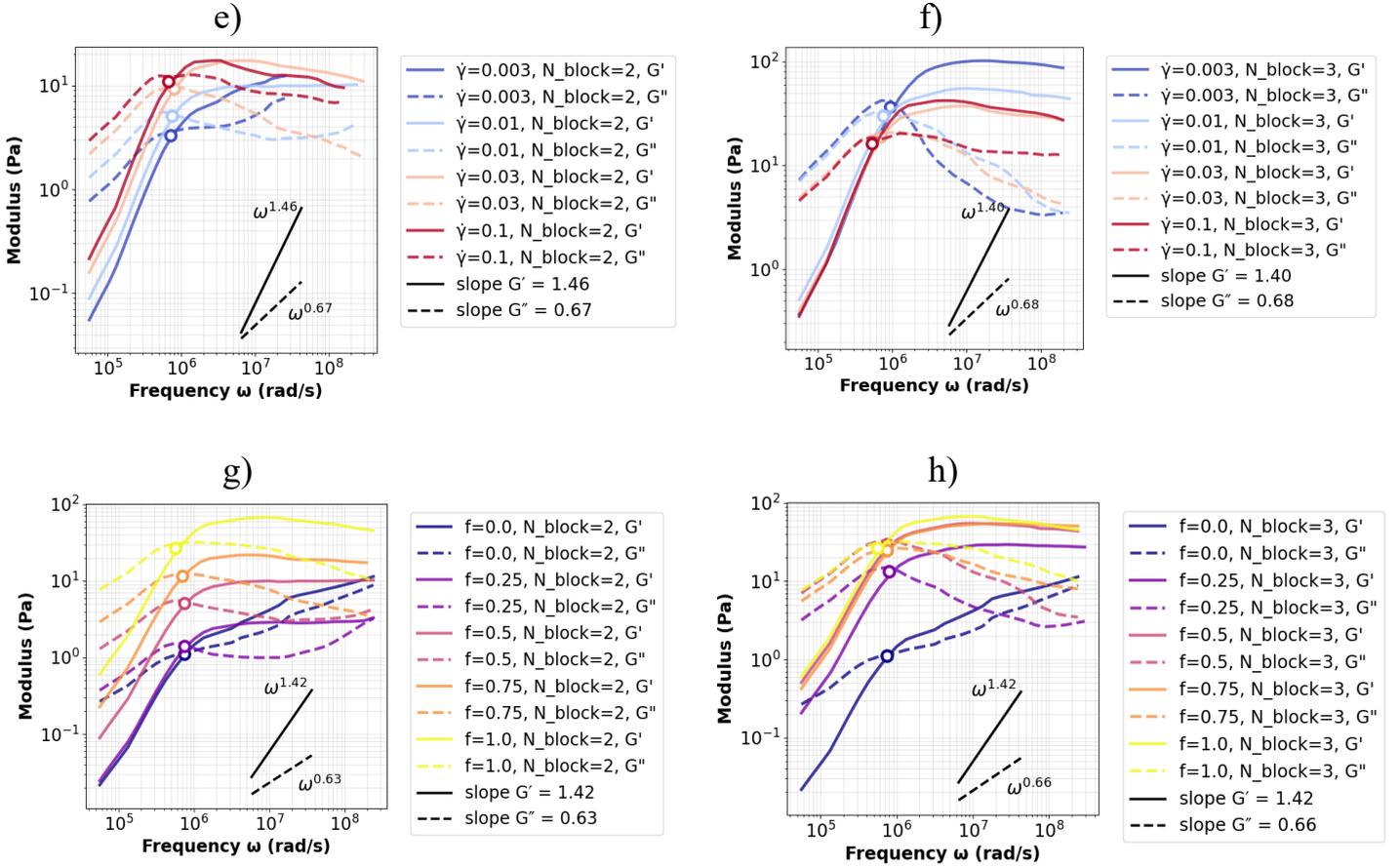

Figure 7: Frequency-dependent storage (G') and loss (G") moduli for diblock and triblock copolymers in dilute (a-d) and semi-dilute (e-h) regimes. Effect of shear rate on a and e) diblock, and b and f) triblock copolymers; and effect of hydrophobic fraction on c and g) diblock, and d and h) triblock copolymers

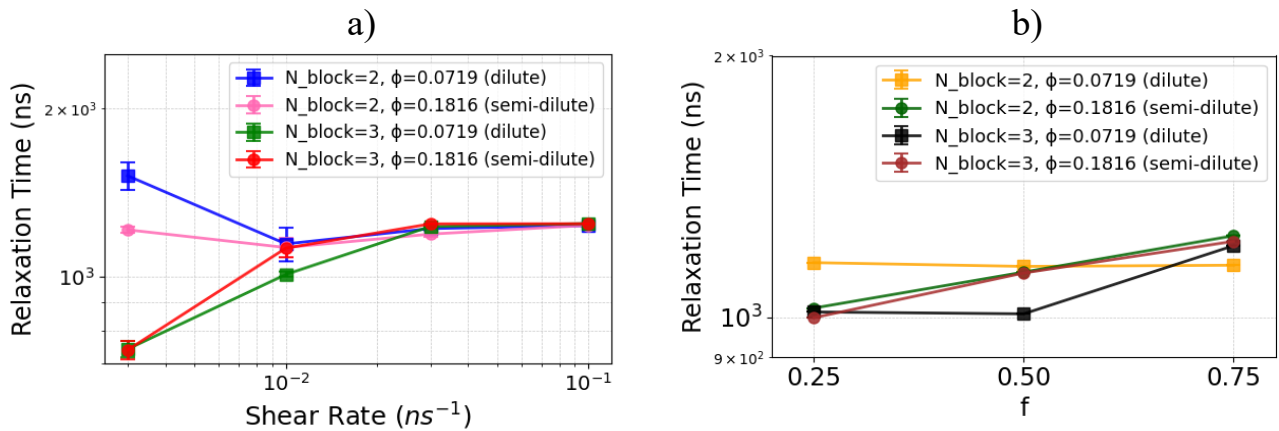

Figure 8: Relaxation time of diblock and triblock copolymers versus a) shear rate at $N$=48, $N_{chain}$=120 and $f$=0.5; and b) hydrophobic fraction in dilute and semi-dilute regimes with $N$=48, $N_{chain}$=120 and $\dot{\gamma}$=0.01 $ns^{-1}$

## 4. Conclusion

This comprehensive computational investigation has elucidated the complex, coupled relationships between molecular architecture, concentration regime, shear flow, and chemical composition in determining both the structural morphology and rheological response of multi-chain amphiphilic block copolymer assemblies. Through systematic Brownian dynamics simulations, a rich morphological landscape encompassing seven distinct aggregate structures has been mapped, and the results connecting nanoscale self-assembly to macroscopic flow behaviour have been reported.

The key findings of this study are summarised as follows:

- **Concentration- and composition-dependent morphological landscape:** In the dilute regime (Ø=0.0719), quiescent conditions produce spherical micelles that transition to cigar-like structures at $\dot{\gamma}$~0.01 $ns^{-1}$, with network fragmentation at higher shear rates; the $f$-dependent progression yields dispersed chains ($f$=0), cigar-like micelles ($f$=0.25), short cylindrical micelles ($f$=0.5), gnarled or worm-like micelles ($f$=0.75), and sheet-like phase-separated structures ($f$=1). In the semi-dilute unentangled regime (Ø=0.1816), shear drives collective network reorganisation toward sheet-like morphologies at moderate rates before fragmentation; the $f$-dependent progression yields cigar-like micelles ($f$=0.25), sheet-like morphologies ($f$=0.5), and necklace micelles ($f$=0.75), with larger phase-separated domains at $f$=1 compared to dilute conditions.

- **Architectural reversal between concentration regimes:** In dilute conditions, triblock architectures achieve higher aspect ratios ($L_1/L_3$~10–11) through inter-micellar bridging networks that collectively align and extend under moderate shear. In semi-dilute conditions, this trend inverts, where diblock systems exhibit higher structural anisotropy owing to greater conformational flexibility enabling sheet formation, while the bridging configurations of triblocks become a topological constraint that resists the two-dimensional flattening required for maximum sheet anisotropy.

- **Concentration-dependent deformation mechanisms:** Dilute systems permit individual micelle elongation with extreme uniaxial anisotropy ($L_1 \gg L_2 \approx L_3$, cigar-like morphologies), while semi-dilute systems undergo collective network reorganisation that distributes stress across interconnected units, promoting biaxial dimensional redistribution ($L_1 \geq L_2 > L_3$, sheet-like and necklace morphologies) and producing systematically lower aspect ratios despite forming larger overall structures.

- **Optimal shear response at moderate rates:** Maximum structural alignment and micellar coalescence occur at $\dot{\gamma} \sim 0.01\ ns^{-1}$ across all systems, with network fragmentation and increasing cluster count occurring at higher shear rates ($\dot{\gamma} > 0.01\ ns^{-1}$), regardless of architecture or concentration regime.

- **Aggregation number scaling confirms concentration-driven regime transition:** Scaling exponents extracted from cluster analysis yield $\alpha=0.833$ in the dilute regime, falling within the theoretical bounds established by Halperin–Alexander (star-like limit $\alpha=0.80$, crew-cut limit $\alpha=1.0$), while the semi-dilute unentangled regime yields $\alpha=1.07$, exceeding the dilute upper bound and confirming that coronal overlap and screening effects at elevated concentration drive stronger aggregation number dependence consistent with the semi-dilute unentangled regime.

- **Rheological architectural inversion:** At equilibrium ($\dot{\gamma}=0$), diblock systems exhibit higher viscosity than triblocks in both concentration regimes due to slower structural relaxation and restricted conformational rearrangement. Under shear, this ordering completely reverses. triblock systems maintain superior viscosity through stable bridging networks that resist flow-induced breakdown, whereas diblock aggregates experience more rapid disruption. All systems exhibit universal shear-thinning behaviour, with the magnitude and rate of viscosity reduction varying systematically with architecture and concentration.

- **Universal non-terminal viscoelastic scaling:** Both $G'$ and $G''$ exhibit power-law frequency scaling deviating markedly from terminal flow behaviour ($G' \sim \omega^2$, $G'' \sim \omega^1$) across all conditions, yielding universal non-terminal exponents of $G' \sim \omega^{1.41 \pm 0.05}$ and $G'' \sim \omega^{0.635 \pm 0.45}$ applicable across both dilute and semi-dilute unentangled regimes, both architectures, and all hydrophobic fractions. This universality arises because scaling is governed by intrinsic micellar relaxation dynamics rather than inter-micellar interactions or concentration-dependent topology. The underlying mechanisms differ between architectures: diblock non-terminal behaviour originates from polydisperse single-chain exchange timescales, while triblock non-terminal behaviour additionally reflects cooperative sequential dissociation of two anchored end blocks, sustaining higher elastic storage at high frequencies without altering the scaling exponents.

- **Architecture- and concentration-dependent micellar relaxation time:** Dilute diblocks exhibit a non-monotonic relaxation time with shear, decreasing to a minimum at moderate shear before recovering at high rates, while dilute triblocks show a monotonically increasing relaxation time as bridging connections are progressively dismantled by flow. At $\dot{\gamma}=0.1\ ns^{-1}$, relaxation times converge across all systems, confirming that flow-induced fragmentation dominates at sufficiently high shear regardless of architecture and concentration regime.
- **Data-driven design**, the insight and design principals emerges from this study can be leveraged with machine learning [36], [37], [38], [39] and optimization algorithms [40], [41], [42], [43] to optimize the self-assembled nanostructures for different purposes and applications.

## CRediT authorship contribution statement

**Ehsan Kamali Ahangar:** Conceptualization, Data curation, Formal analysis, Investigation, Methodology, Software, Validation, Visualization, Writing – original draft. **Dominic Robe:** Software, Co-supervision, Writing – review and editing. **Elnaz Hajizadeh:** Project administration, Resources, Supervision, Writing – review and editing.

## Declaration of competing interest

The authors declare no conflict of interest.

## Acknowledgements

The first author acknowledges with gratitude the financial support provided through the Melbourne Research Scholarship (MRS) from the University of Melbourne. Computational resources and technical assistance provided by Research Computing Services at the University of Melbourne are also gratefully acknowledged.

## Supporting information

The Supporting Information is available here.

Figure S1: Eight-stage simulation workflow for multi-chain block copolymer systems. Stage 1: Parameter space definition using *signac* framework; Stage 2: Automated generation of input files and initial configurations based on specified parameters; Stage 3: Energy minimization and four-step thermal annealing to prevent kinetic trapping; Stage 4: Brownian Dynamics simulations under both shear and quiescent conditions, capturing conservative, stochastic, and friction forces; Stage 5: Production runs with real-time monitoring over 2200 *ns*; Stage 6: Post-processing analysis extracting target properties using LAMMPS output and Python scripts; Stage 7: Morphology-dependent characterization based on hydrophobic fraction, shear rate, chain architecture, and volume fraction using specialized Python libraries; Stage 8: Statistical analysis aggregating results across independent replicas.